\documentclass[sigconf]{acmart}

\usepackage{latexsym}
\usepackage[utf8]{inputenc}
\usepackage{graphicx}
\usepackage{comment}
\usepackage{url}
\usepackage{csquotes}
\usepackage{dirtytalk}
\usepackage{hyperref}
\newcommand{\changes}[1]{\textcolor{black}{#1}}
\AtBeginDocument{%
 \providecommand\BibTeX{{%
 \normalfont B\kern-0.5em{\scshape i\kern-0.25em b}\kern-0.8em\TeX}}}

\copyrightyear{2022}
\acmYear{2022}
\setcopyright{rightsretained}
\acmConference[CHI '22]{CHI Conference on Human Factors in Computing
Systems}{April 29-May 5, 2022}{New Orleans, LA, USA}
\acmBooktitle{CHI Conference on Human Factors in Computing Systems
(CHI '22), April 29-May 5, 2022, New Orleans, LA, USA}
\acmDOI{10.1145/3491102.3517446}
\acmISBN{978-1-4503-9157-3/22/04}

\begin{document}

\title[Are Deepfakes Concerning?]{Are Deepfakes Concerning? Analyzing Conversations of Deepfakes on Reddit and Exploring Societal Implications}

\author{Dilrukshi Gamage}
\affiliation{%
  \institution{Tokyo Institute of Technology}
  \city{Tokyo}
  \country{Japan}}
\email{dilrukshi.gamage@acm.org}

\author{Piyush Ghasiya}
\affiliation{%
  \institution{Tokyo Institute of Technology}
  \city{Tokyo}
  \country{Japan}}
\email{ghasiya.p.aa@m.titech.ac.jp}

\author{Vamshi Krishna Bonagiri}
\affiliation{%
  \institution{International Institute of Information Technology}
  \city{Hyderabad}
  \country{India}}
\email{vamshi.b@research.iiit.ac.in}

\author{Mark E Whiting}
\affiliation{%
\institution{ The Wharton School, University of Pennsylvania}
\city{Philadelphia}
\state{Pennsylvania}
\country{USA}}
\email{markew@seas.upenn.edu}

\author{Kazutoshi Sasahara}
\affiliation{%
  \institution{Tokyo Institute of Technology}
  \city{Tokyo}
  \country{Japan}}
\email{sasahara.k.aa@m.titech.ac.jp }


\renewcommand{\shortauthors}{Gamage, et al.}

\begin{abstract}
Deepfakes are synthetic content generated using advanced deep learning and AI technologies. The advancement of technology has created opportunities for anyone to create and share deepfakes much easier. This may lead to societal concerns based on how communities engage with it. However, there is limited research available to understand how communities perceive deepfakes. We examined deepfake conversations on Reddit from 2018 to 2021---including major topics and their temporal changes as well as implications of these conversations. Using a mixed-method approach---topic modeling and qualitative coding, we found 6,638 posts and 86,425 comments discussing concerns of the believable nature of deepfakes and how platforms moderate them. We also found Reddit conversations to be pro-deepfake and building a community that supports creating and sharing deepfake artifacts and building a marketplace regardless of the consequences. Possible implications derived from qualitative codes indicate that deepfake conversations raise societal concerns. We propose that there are implications for Human Computer Interaction (HCI) to mitigate the harm created from deepfakes.
    
\end{abstract}

\ccsdesc[500]{Human-centered computing~Empirical studies in collaborative and social computing}
\ccsdesc[500]{Human-centered computing~Social content sharing}
\ccsdesc[300]{Human-centered computing~Social media}
\ccsdesc[500]{Human-centered computing~Social content sharing}
\ccsdesc[300]{Human-centered computing~Social engineering (social sciences)}
\ccsdesc[100]{Human-centered computing~HCI theory, concepts and models}


\keywords{deepfake, societal implication, content analysis, topic modeling}

\begin{teaserfigure}
 \includegraphics[width=\textwidth]{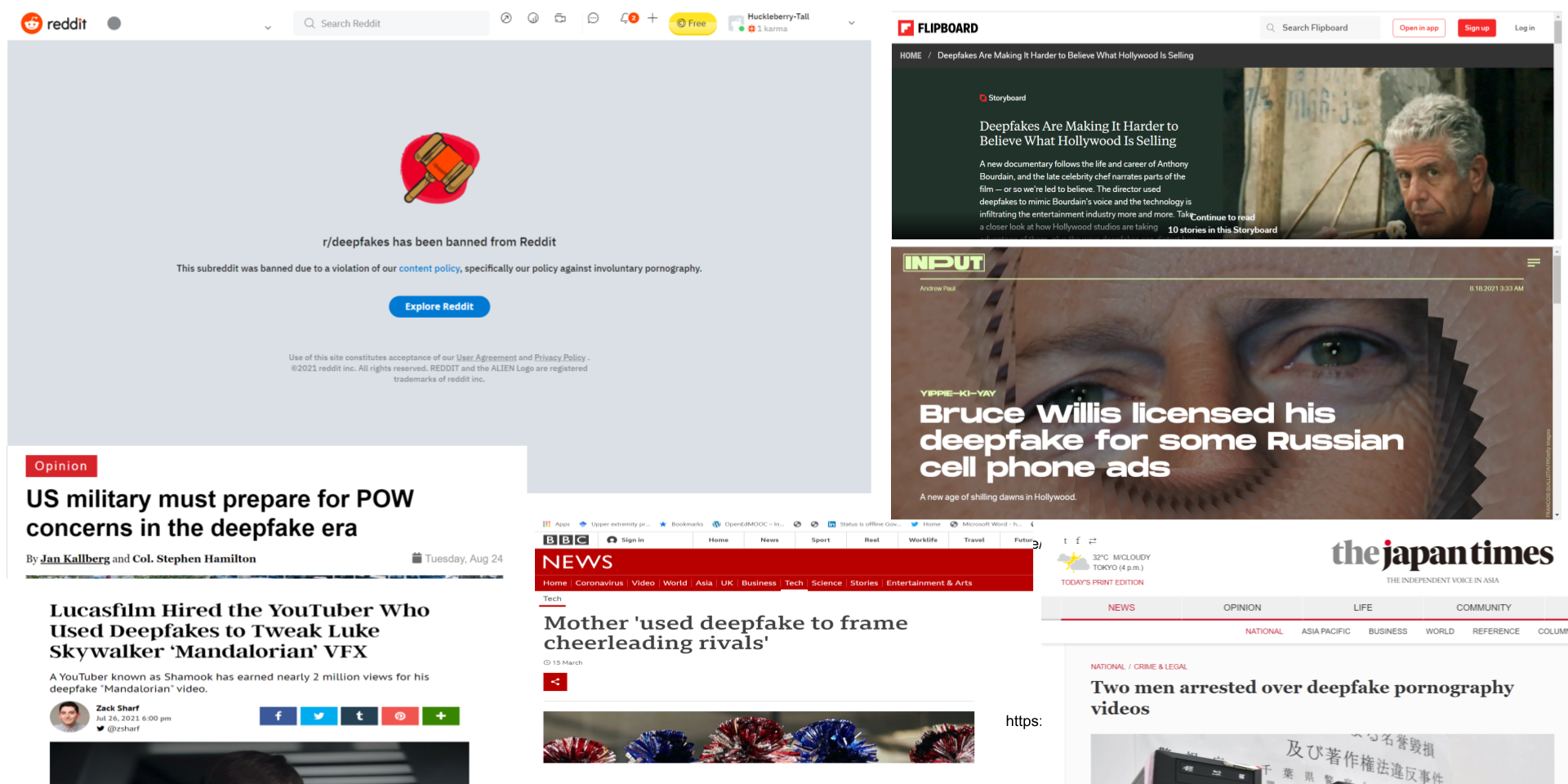}
 \caption{Some stories and events behind deepfakes obtained from Reddit web (2018) and BBC, Japan Times, Indiewire, and Flipboard from March to August 2021}
\label{fig:story}
\end{teaserfigure}

\maketitle

\section{Introduction}
Deepfakes are the application of deep learning methods to generate fake content---usually video, images, audio or text. The quality of content created by deepfake technology is improving, making it is indistinguishable from real content. This involves transferring one's voice, video, or image to another to reflect the original person, mostly done without consent. However, deepfakes are deeply contentious---on one hand, they undermine our trust in content, on the other hand, they open up new creative opportunities. Deepfakes impact the information consumption experience and have already created negative consequences in society. Although most deepfakes are considered harmful, there has been lengthy debate about ethical implication~\cite{de2021distinct}. One of the latest much debatable incidents was in the use of a deepfake voice in the documentary film ``Roadrunner'', a film about a celebrity chef Anthony Bourdain, who ended his own life due to depression. The film director incorporated AI synthetic voice to show Bourdain speaking for real as a part of the scene. However, such a voice cut was never originated or consented to by the late chef himself. This incident brought much discussion and debate on the ethical implications of the use of deepfake to create fake media~\cite{dfethics2021}.

Deepfakes can harm viewers by deceiving or intimidating them. They can harm victims by causing reputational damage, and beyond that, they impact society by undermining societal values, i,e., trust in institutions and individuals. The term ``deepfake'' was seen in 2017 when a group of Reddit users was responsible for creating synthetic celebrity pornographic videos using AI technology~\cite{maddocks2020deepfake}. Since then such non-consensual synthetic media has been impacted society. It can be harmful to democracy by disseminating political disinformation~\cite{vaccari2020deepfakes}, undermining national and international security by sharing nonconsensual synthetic content~\cite{sayler2020deep}, and complicating law enforcement. Further, it can be harmful to the entertainment industry by disseminating fake celebrity pornography~\cite{harris2018deepfakes}. However, apart from such multi-level impacts, due to the advancement and democratization of AI tools, domestic level daily usage of such technologies is emerging and increasingly affecting our society (see some stories of Deepfakes in Fig~\ref{fig:story}~\cite{japanarrest,stupp2019fraudsters,Guardian_2020}). 
Although the deep learning that produces deepfake are versatile and could be useful in revolutionizing various industries such as online courses and customer service, the negative incidents collectively raise concerns about the societal problems emerging from them. Currently, much attention from research, academia, and industry are being directed toward ML, AI, and deep learning for the creation~\cite{yadav2019deepfake, caldelli2021optical} and detection of deepfakes~\cite{maksutov2020methods, rana2020deepfakestack}. However, how humans perceive or interact with deepfake content or how deepfake is used and its societal implications are less examined. In particular, understanding the communities that create these deepfakes, and how these communities in platforms discuss the deepfakes remain unexplored.

The Reddit subgroup that created the first set of deepfakes impacted the society as it involved nonconsensual pornographic videos. Although this sub Reddit group was banned from the platform, we found many other discussions about deepfakes on Reddit. These discussions may or may not lead to harmful implications since these conversations have not been critically examined. However, emerging social spaces that discuss deepfakes need extensive interrogation and comprehension of how this phenomenon is situated in the everyday public conversation and what directions these conversations are headed. Currently, myriad of research discussing the societal implications of deepfake has been centered upon critical reviews. However, empirical research is needed to identify the human-centric social process of deepfakes, and the implications should be investigated more by making an interdisciplinary effort since it is important to identify social measures to counter with deepfakes.

In this work, we couple methods from computational social science and human-computer interactions to investigate the current landscape of deepfake discussions within the society. We observed and analyzed a wide variety of discussions about deepfake in Reddit. Since Reddit communities are anonymous in nature, we believe that their discussions may provide valuable insights into the implications of deepfakes that mainstream social media platforms may impede. Therefore, to explore perspectives of deepfakes in the community, we addressed the following questions:
\begin{itemize}
    \item 
\textit{ RQ1:} What type of conversations lead in Reddit communities about deepfakes and how do these conversations change over time?

\item 
  \textit{ RQ2:} What are the societal implications visible in these conversations?
\end{itemize}

To answer these questions, we obtained Reddit conversations related to deepfakes from 2018 to 2021 and followed a mixed-method approach. First, we conducted a large-scale quantitative analysis from a computational social science perspective to understand what lies beneath the bigger picture of the conversations. Second, an in-depth qualitative analysis was conducted to draw deeper insights into the conversation and its possible implications. 
Topic modeling in natural language processing (NLP) was employed to understand major topics and qualitative coding was used to map the key themes. We found that many conversations lead to dominant discussion topics such as ---political discussions, pornography discussions, concerns about the ``realness'' of deepfake and discussions about how platforms should act to mitigate the harm caused by it. However, we also found that pro-deepfake behaviors in Reddit where users are highly supportive to create deepfakes and even some users provide deepfake as a service and monetize deepfake creations. The qualitative codes indicate that deepfake conversations are leading to concerning implications and may be harmful to the society at large in the future.

Early identifications of the possible effects of deepfakes help the Human Computer Interaction (HCI) research communities to understand the needs, and articulate the designs, and interactive technologies that reduce plausible societal harms. This research contributes to HCI by identifying implications while reflecting empirical evidence on the key topics of deepfake conversations. We further address how the nature of conversation changes over time and the possible implications from these topics whereby future researchers could work on directions for mitigating the harm from deepfakes. We as HCI researchers are the gateway to understanding human behavior and it is important to design solutions centered upon the user. In this case, we see constant movements enforcing deepfake detection methods from computer vision and AI researchers, yet we believe this could be benefited and need attention from social scientists and HCI researchers to examine it as a community problem. Combinations of disciplines need closure examining any solution and this research has just opened a Pandora's box. Our dataset, computational scripts and qualitative codes with descriptive snippets will be publicly available to reuse and replicate the studies---at \url{https://osf.io/dg2wh/}.

\section{Background and review of the literature }
In this section, we briefly discuss the background of the current techniques, the types of deepfakes found in society and the research landscape. We also provide a review of the research centered around Reddit communities. 

\subsection{Deepfake technology} 
Deepfake is synthetic media where one person’s voice, image, or video is replaced by someone else, thus appearing to be the original person speaking or their photo. The technology for deepfake leverages ML and AI techniques. Primarily, deep neural network models are used, such as generative adversarial networks (GAN). These generate visual and audio content with a high potential to deceive. Deepfakes predominantly represent research by academics and industries in the specific fields of computer vision and ML/AI in computer science. They mostly focus on projects for improving deepfake techniques~\cite{bregler1997video, suwajanakorn2017synthesizing, thies2016face2face, chan2019everybody} and methods for detecting deepfakes~\cite{zheng2019survey, li2020face, li2020fighting, pu2020noisescope,agarwal2021detecting}. A recent survey on deepfake creation and detection techniques predicts that deepfakes can be weaponized for the monetization of its products and services~\cite{mirsky2021creation}. Since the tools are becoming more accessible and efficient, it seems natural that malicious users will find ways to use the technology for profit. As a result, researchers expect to see an increase in deepfake phishing attacks and scams targeting both companies and individuals~\cite{mirsky2021creation}.

\subsection{Types of deepfakes in society}

The deepfake phenomenon must be positioned within the sphere of society and should be examined under the lens of the contexts of its uses (or potential uses), along with its effects. Currently, researchers indicate that deepfakes will have implications in law and regulations, politics, media production, media representation, media audiences, and overarching interactions in media and society~\cite{karnouskos2020artificial,mirsky2021creation}. With the commercial development of AI applications such as FakeApp (\url{https://www.faceapp.com/}), Deepfake Web (\url{https://deepfakesweb.com}) and Zao (\url{https://zaodownload.com/}) or audio deepfake apps such as Resemble (\url{https://www.resemble.ai}), and Descript (\url{https://www.descript.com/overdub}) as well as with the use of synthesized text~\cite{jaderberg2014synthetic}, it is evident that attacks involving human reenactments and replacement are emerging.
Cyber security reports in 2019 predict 96\% of all deepfakes to be pornographic~\cite{analytica2019deepfakes} and other forms of harmful outcomes~\cite{oxford2019deepfakes, zeng2019preparing,perez2021deepfakes}. In addition, research reveals alarming concerns over developments in deepfake child pornography~\cite{eelmaa2021sexualization} or applications such as DeepNude where algorithms remove clothes from images of women~\cite{greengard2019will}.

Furthermore, the current platforms' business model is content-driven and use incentives to increase the number of views and likes. Platforms such as Facebook, Twitter, YouTube, TikTok, and Reddit are all competing for users' attention ~\cite{schirch2021techtonic}. Due to significant monetary incentives, content creators are pushed into presenting dramatic or shocking videos, where potential deepfakes may draw the attention of the public~\cite{langguth2021don}. Although some research suggests that the increased proliferation of deepfake videos will eventually make people more aware that they should not always believe what they see~\cite{langguth2021don}, much empirical research is essential to understanding the impact and estimating the potential damage where interventions are needed before deepfake causes harm.

\subsection{Social centric research for deepfake}

This section provides a critical review of research on the societal implications of deepfakes. Although limited research can be found providing social evidence of how people react or what they understand when they see deepfakes, a study conducted in 2020 using a sample of 2,005 UK citizens provides the first evidence of the deceptiveness of deepfakes~\cite{vaccari2020deepfakes}. In published articles discussing deepfake societal implications, the authors have brought many anecdotal shreds of evidence leading to the prospect of mass production and diffusion of deepfakes by malicious actors and how it could present extremely serious challenges to society~\cite{fallis2020epistemic}. Since the social perspectives are based on ``seeing is believing'' or ``a picture is worth a thousands of words'', images and videos have much stronger persuasive power than text; thus citizens have comparatively weak defenses against visual deception of this kind~\cite{newman2015truthiness}. Particularly the research by ~\cite{vaccari2020deepfakes} investigated how people can be deceived by political deepfake, and it turns out that society is more likely to feel uncertain than to be completely misled by deepfakes. Yet the resulting uncertainty, in turn, reduces trust in the news on social media. The societal implications of deepfakes can be easily grounded when we understand how people interact with such technology. 

Overall, considerable research has highlighted that deepfakes are likely to attack in the spheres of political disinformation and pornography. One of the latest research studies indicates the likelihood of sharing deepfakes based on individuals' cognitive ability, political interest and network size~\cite{ahmed2021inadvertently,cochran2021deepfakes}. 
Although deepfakes are relatively new and risks have yet to be seen, many reports show that deepfakes have the potential to influence multiple sectors of society such as the financial sector which relies on information, politics, national and international security~\cite{brooks2021popular}. They may even threaten private lives by revenge porn and faked evidence used in legal cases resulting in distress and suffering~\cite{brooks2021popular}. 
Although many in the field of communications, media, and psychology try to understand the impact of deepfakes, areas such as human-computer interaction and human-centric designs have sparsely been covered in exploring deepfake effects on society. To highlight, we found only nine research articles from the search term ``deepfake'' in the prominent conference on Computer Human Interaction conference (CHI)(accessed May 2021). Of the nine articles, only two articles involved perceptions of deepfakes' impact on society. Of those two, one research article conducted experiments focusing on psychologically measurable signals which could be used to design tools to detect deepfake images~\cite{wohler2021towards}. The other study more importantly addressed how human eyes detect deepfake while providing implications of a designed training for distinguishing deepfakes, especially in low literate society ~\cite{tahir2021seeing}.  
By far one of the most factual evidence of widely known deepfake effect---revenge porn is described in an article from CHItaly. Their research provides an expert analysis of the process of reporting revenge porn abuse in selected content sharing platforms such as image and video hosting websites and pornographic sites. In a search for a way to report abuse by revenge pornography using deepfakes, they found serious gaps in technology designs to facilitate such reporting. This research provides a view of current practices and potential issues in the procedures designed by providers for reporting these abuses~\cite{de2021reporting}. With all of the research, it is evident that deepfakes could damage individuals and society. Prevention and detection may be one way of looking at it. The design of digital technologies will play an important role in reacting to any social issue cause by deepfakes and interdisciplinary approaches must be taken to address these issues. 

In this research, we examine conversations on Reddit related to deepfakes and identify emerging threats while proposing solutions integrating changes in technological and societal behavior.

\subsection{Analysis of Reddit communities}
Historically, studies conducted on Reddit have dealt with sensitive topics~\cite{sowles2018content,wang2015examination,maxwell2020short}, mainly to obtain deep insights into topics that individuals may not discuss in open public spaces where they can be identified. We highlight recent related research to examine Reddit users' views and attitudes about the sexualization of minors in deepfakes and ``hentai'' which is known for overtly sexual representations (often of fictional children). Based on a large dataset of Reddit comments ($N=13,293$), the analysis captured five major themes regarding the discussions of sexualization of minors: illegality, art, promoting pedophilia, and an increase in offensive and general harmfulness. The study provides information that could be useful in designing prevention and awareness efforts that address the sexualization of minors and virtual child sexual abuse material~\cite{eelmaa2021sexualization}. Similarly, another research analyzed domestic abuse discourse on Reddit~\cite{schrading2015analysis}. The authors have developed a classifier to detect discussions about domestic abuse, and the study provides insight into the dynamics of abusive relationships.~\cite{schrading2015analysis}. This study has followed a similar design found in an analysis of mental health where researchers used sub Reddits on the mental health topic and found differences in discourse~\cite{balani2015detecting}. This shows that observing particular sub Reddits and qualitative coding for a sample of posts in selected sub Reddits are steps commonly followed by researchers. In our study, however, we focused on deepfake posts and comments but not on a particular sub Reddits, because we were interested in wider perspectives on deepfakes. 

Apart from qualitative analysis, quantitative approaches are also used for Reddit analysis. One study conducted topic modeling and hierarchical clustering to obtain both global topics and local word semantic clusters to understand factors affecting weight loss. The authors employed a regression model to determine the association between weight loss and word semantic clusters, and online interactions ~\cite{liu2020understanding}. 
Although any social media platform's user-generated content and activity offer a unique opportunity to understand how individuals and social groups interact toward certain phenomena, we specifically used Reddit to understand the dynamics of social interactions centered around deepfakes. Since Reddit is the birthplace of deepfakes, and where certain sub Reddits responsible for deepfake were supposedly banned~\cite{maddocks2020deepfake}, we observe emerging discussions on deepfakes in many other sub Reddits (i.e., r/deepfakes). Unlike some studies that use one or more particular sub Reddits to understand communities, we observed a wide range of sub Reddits such as technology, movies, finance, etc. are discussing deepfakes as trending posts. Although we found a study on deepfake discourse using Twitter data~\cite{perez2021deepfakes}, which discusses the characteristics of users of deepfake content sharing and their demographic details, our study digs insight into the dynamics of discourse in online communities on Reddit where these anonymous users could offer valuable insights in the context, helping to understand the emerging conversations toward deepfakes. Analyzing these conversations is likely to open the Pandora's box and question the effects that may be created by future communities that might pose a threat to society. To date, Reddit communities have not been used to understand deepfake's societal implications, yet researchers emphasize the important need for such an analysis~\cite{hancock2021social}. 

\begin{figure*}
 \centering
 \includegraphics[width=\linewidth]{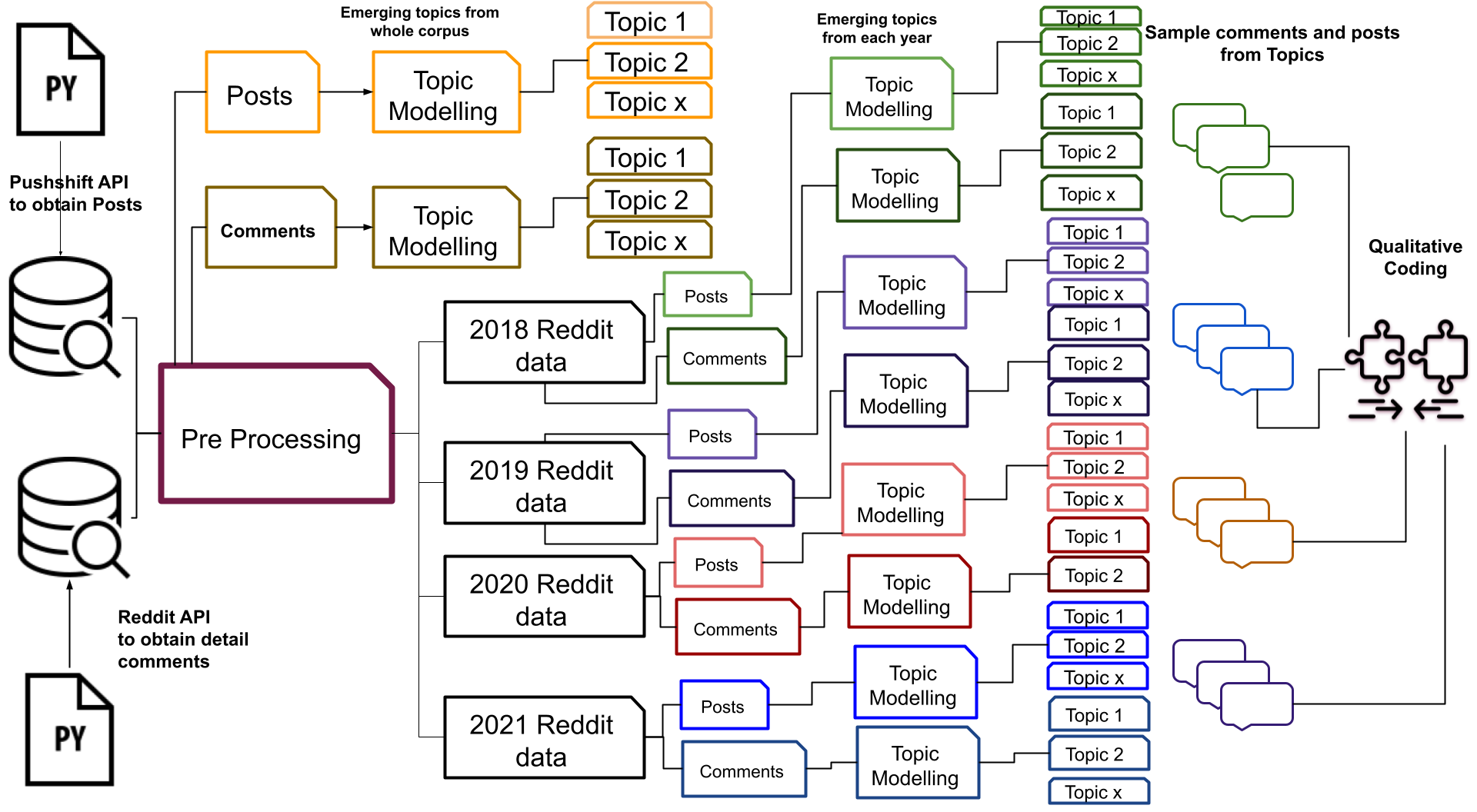}
 \caption{Process of obtaining data, pre-processing, topic modeling for data in 2018 to 2021 as a whole and for each year. At the final phase, qualitative coding was conducted for the sample comments and posts obtained from each topic.}
\label{fig:process}
\end{figure*}

\section{Methods}
We have followed a mixed-method approach intersecting with the HCI research tradition and computational social science where similar approaches are found in Reddit analysis~\cite{kimo2020comparative, thukral2018analyzing}, especially the qualitative aspects~\cite{otiono2019understanding} and quantitative approaches~\cite{chandrasekharan2018internet}. Data were collected from Reddit between 2018 and 2021. We used the Pushshift Reddit API (\url{https://github.com/pushshift/api}), passing a query for the search term ``deepfake'' to obtain all the posts and Reddit API to obtain all replies to the posts as comments. In return, the script provided all posts (original submissions  to Reddit as posts), comments (replies to the posts), dates, names of sub Reddits, and usernames of the comments and posts. 

\subsection{Ethical considerations}
Our universities do not require ethical clearance to obtain and analyze publicly available data that does not involve human subjects. At the same time, the CHI and CSCW communities are in the early stages of discussing and forming guidelines for conducting research using public social media~\cite{otiono2019understanding}. However, we acknowledge the ethical concerns associated with our methods. We obtained only publicly available Reddit data and did not reach any private sub Reddit groups. We understand that contributors to public Reddit groups may not be aware that their conversations are used for academic research and we have not explicitly obtained their consent for this research. We also understand that user names of Reddit are anonymous or users are independent of their real names, yet Reddit does not guarantee anonymity. Therefore, in this study, we did not focused on specific user perspectives or specific sub Reddit or specific moderators in the groups. Instead, our analysis focuses on large-scale data, focusing on discussion topics, user perspectives and drawing insights from data as a collection. Our motivation is driven to make a safer environment and mitigate the harms arise from deepfake technology. We believe that early understanding and perceptions on deepfakes will reflect the impacts and benefit society as it will provide design directions for socio-technical tools to mitigate the risks. For reproducibility, our scripts and data (without usernames) are available in the OSF repository.

\subsection{Data and Analysis}
The ``deepfake'' search query yielded a large data set with 86,425 comments distributed among 6,638 posts, between January 1, 2018 and August 21, in 2021. As previously mentioned, posts are the original submissions and comments are the replies to the posts. We analyzed these separately to determine if replies are in agreement with posts by discussing similar topics.

To answer RQ1: What are the main types of conversations in Reddit communities about deepfakes and how have these conversations change over time?---we relied on a method commonly used in computational social sciences. For the first part, we leaned on natural language processing (NLP) method of Topic modeling. This is an unsupervised approach to recognizing topics by detecting patterns of word occurrences. It helps to understand and summarize large collections of textual information and discover latent topical patterns present across the collections. For the second part of the question, the temporal analysis of posts and comments was used to reflect the shift in the types of posts and comments, and the direction of leading topics. 

To answer the RQ2: What are the societal implications visible in these conversations?---we relied on a qualitative method --- open coding used in HCI. Figure~\ref{fig:process} depicts the overall procedures.

\subsubsection {Text Preprocessing}
Before performing the topic modeling, it was essential to prepare and clean the data. We used Python NLTK libraries for text preprocessing~\cite{bird2009natural} and followed the steps commonly used in topic modeling. These involve expanding the shortened versions of words (don't, or I'd..etc,) removing links, tokenization which converts each word to a token, lemmatization which convert each token into their root word, removing stop words (a, an, the, etc), removing  special characters, converting all characters into standardized ASCII characters and making all words into lowercase.

\subsubsection{Topic Modeling} 
After preprocessing the data, we performed the topic modeling to discover the abstract ``topics'' that occur in both posts and comments. There are several topic modeling  methods such as Latent Dirichlet Allocation (LDA), Latent Semantic Analysis (LSA), Non-Negative Matrix Factorization (NMF). We used the NMF model, since our data contained short texts rather than long documents. However, to produce the best topic model, prior knowledge of the optimal number of topics present in the data is essential. This (optimal number of topics) is one of the main challenges in the topic modeling approach. Usually, it is done manually (eyeballing) by running the model with a different number of topics and selecting the model that produces the most interpretable topics. If Topic Coherence-Word2Vec (TC-W2V) metric is used with NMF, it can find the optimal number of topics automatically~\cite{o2015analysis}. This metric measures the coherence between words assigned to a topic. For example, how semantically close are the words that describe a topic. We trained the word2vec model~\cite{mikolov2013distributed} on our corpus, which would organize the words in an n-dimensional space where semantically similar words are close to each other. The TC-W2V for a topic will be the average similarity between all pairs of the top $n$ words describing the topic. We then trained the NMF model for different values of the topic ($k=1$ to 19) and calculated mean topic coherence across all the topics. The $k$ with the highest mean topic coherence is used to train the final NMF model. For the above implementation, we used the scikit-learn python library and built an NMF-based topic 
First, we performed topic modeling on the entire corpus grouped by ``posts'' and ``comments''. This included all the posts and comments between 2018 to 2021. This provided an overall view on what users post and comment. We also measured the weights of the terms in each topic to understand which terms were highly representative of the topic. Next, we chunked the corpus group by each year and performed topic modeling. This provided an overview of hidden thematic information of engagement in the form of posts and comments each year. This helped us to understand if topics in deepfake discussions were changing over time.

\subsubsection{Temporal Analysis}
In the temporal analysis, we summarized the number of posts and the number of comments distributed in each month of the year for all four years (2018-2021). We also provided a summary of posts, comments and number of sub Reddits in each year which provided an overview of engagement frequency over time.

\subsubsection{Qualitative coding}
We used open coding for mapping the discussed topics with possible implications to the society. We obtained 10 posts and comments from each topic from each year as the sample data. Since the total numbers of posts and comments were too large to analyze by human coders, 10 per topic provided us a considerable sample to analyze qualitatively. The topics (semantically coherent set of words) and the sample posts provided the context of the topics to generate qualitative codes which can be derived for possible implications.

All the topics and sample posts from each topic examined by two researchers who independently identified open codes. These codes were derived based on possible implications that could occur from these conversations. Next, they and two other researchers collectively discussed the open codes. They compared the codes, resolved any disagreements, and finalized the key themes following the inductive process categorizing into groups with similar outcomes. The outcome of these themes resonate with implications. Of the four researchers, two had extensive experience in qualitative data analysis and all four had a sense of the gathered data, as well as an understanding of the context of deepfakes, and the practical knowledge to determine the capabilities of deepfake technology. 

\section{Results}
\subsection{RQ1-First part: What are the main type of conversations in Reddit communities about deepfake?}

\subsubsection{Overall topics from posts and comments from all four years}
Topic modeling was conducted for posts and comments separately. The corpus had 86,425 comments and 6,638 posts. According to the NMF topic coherence graph, there were two dominant topics for both posts and comments. 
The two topics generated out of posts had a coherence measurement of nearly 0.980, while the two topics in the comments had a coherence of slightly above 0.560.

From the \textit{posts}, out of two topics, Topic 1 indicates that major discussions are centered on videos created using deepfakes, specifically related to former US President Trump. This indicates that one of the major posted topics in deepfakes is related to politics and political figures. Topic 2 reveals that the highest weighted coherent word is ``porn'', followed by ``video'' and ``nude''. This reflects that many Reddit submissions are focus on pornographic videos, specifically igniting discussions of fake nudity and the technologies involve in creating such videos. The \textit{comments} reflect the reply behaviors to the posts. There were two topics generated from replies to comments. Topic 1 shows that many users expressed concerns over the ``fakeness'' and that discussions were centered around how to identify such videos of deepfakes. Topic 2, showed a tendency for people posting further links relating to posts and platforms taking actions against deepfakes. Some of these posts may have been deleted by bots or moderators closely monitoring the comments where users were requesting to re-post links from others. 

Next, we obtained the results of topic modeling constructed from the corpus of each year. A summary of all years, comments and posts across the sub Reddits generated a number of topics and the glimpses of the growth of conversations over time with wider sub Reddit groups are described. A summary of projection relevant to each topic number is depicted in Fig.~\ref{fig:compare}. The appendix of this manuscript provides details of the semantically coherent weight of the word distributions reflecting each topic from posts and comments of all years.

\subsubsection{Conversations during 2018}
The dataset from 2018 contained 269 posts across 165 sub Reddits and 3828 comments. We conducted a similar analysis for \textit{posts} and \textit{comments} in 2018. Compared to other years, 2018 had the fewest conversations and we believe this was the result of Reddit's ban of the sub Reddit "s/deepfakes". Reddit subsequently updated their site-wide rules against involuntary pornography and sexual or suggestive content involving minors~\cite{burkell2019nothing}. Examining the total number of posts from 2018, we found only one key topic. This reflects that during that time, the conversation centered more around deepfake technology, the news it had created based on the ban, pornography, and tech news about the new regulations. 

\begin{figure*}
 \centering
 \includegraphics[width=\linewidth]{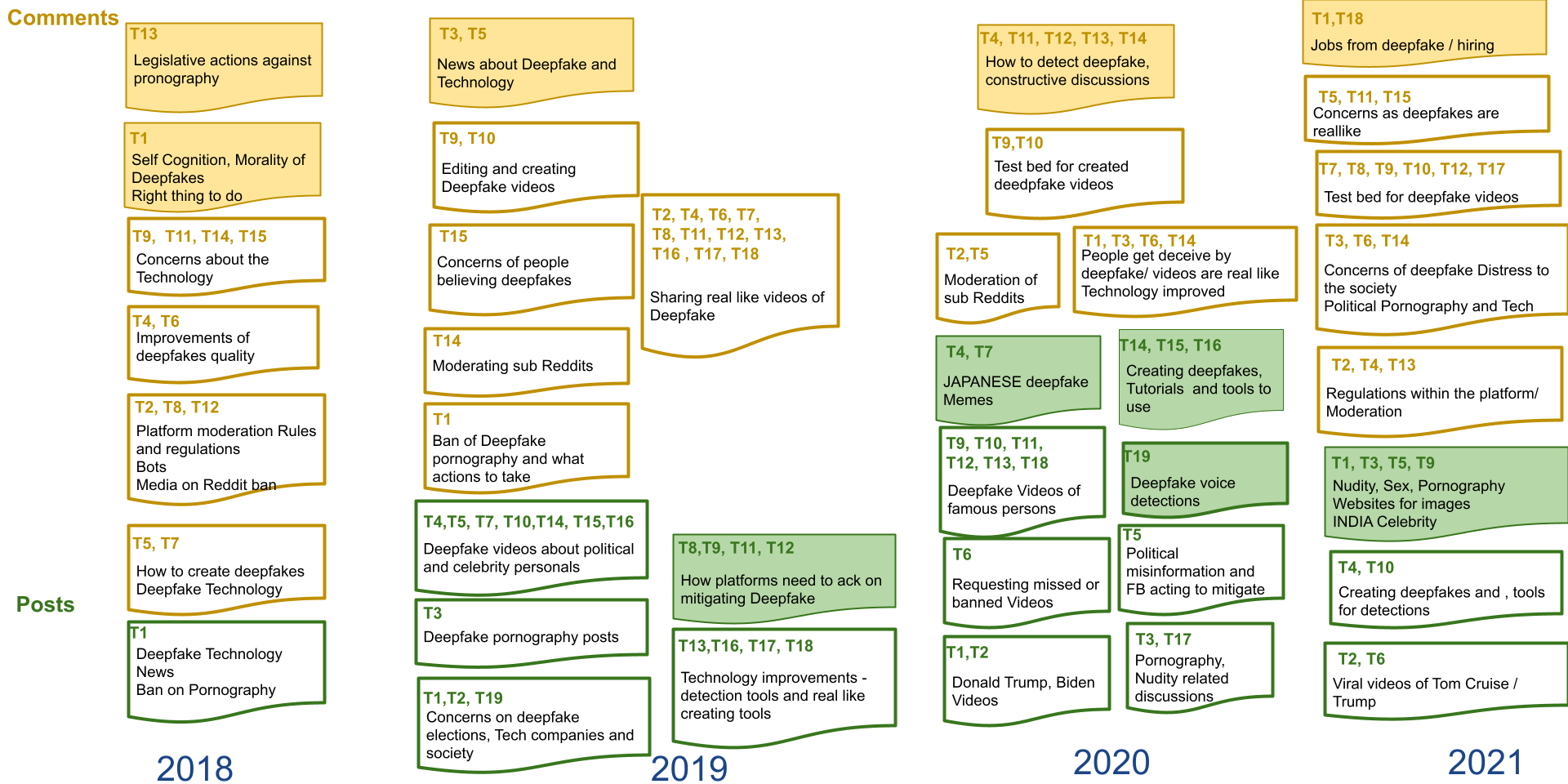}
 \caption{The NMF model resulted number of topics and topic numbers are manually grouped together based on similarity to compare overall discussion from posts and comments over the years. Green outlines are the posts and orange outlines are comments. Comments and posts unique to each year are filled with color.}
\label{fig:compare}
\end{figure*}

Figure~\ref{fig:compare} displays the key topics from comments and posts that resulted from our scripts in 2018. The key topics were grouped based on semantic similarity with topic numbers. Compared to the posts in 2018, there was much engagement in distributed comments. We found 15 topics generated from comments. According to topic distribution, the top ten keywords from the topics and the raw data related to topics, we found that Topic 1 was about community discussions regarding self-cognition, the morality of deepfakes or proactive discussions---conversations on the right thing to do. Topics 2, 8, and 12 are related to bots, shared links, and moderation, new rules, Reddit bans and related media on the announcement by Reddit. Topics 4 and 6 collectively reflect discussions on the improvements to deepfakes as the ability to deceive more people increased as well as the possibility of gaining wider attention and receiving media coverage. Topics 5 and 7 together reflect conversations on technologies for deepfake, Faceswap, and how to create quality deepfake videos or images. Topics 8 and 12 lean more towards platform's moderations. However, topic 13 stands out on its own---it brought key conversations on the legislative actions on revenge porn, celebrity porn, and discussions about consents of the victims. Topics 3 and 10 did not provide meaningful conversations. Topic 9, 11, 14, and 15 together provided deep concerns on the future of technology.

\subsubsection{Conversations during 2019}
Compared to 2018, 2019 had an increased number of conversations (22,773) and posts (538). Based on the topic distribution of the posts, we received 19 topics. Post represented more targeted conversations than in 2018. Interestingly, there were fewer posts on pornography as Topic 3 is the only topic posted on that subject. Topics 1, 2, and 19 describe posts leaning towards concerns about deepfake videos, especially the experts' views on threats for the elections, and the scariness of the future behaviors of tech companies to the society. Topics 4, 5, 7, 10, 14, 15, and 16 collectively provide posts about deepfake videos on political leaders such as Donald Trump, Vladimir Putin, and Barack Obama and celebrity deepfakes including Tom Cruise, Keanu Reeves, Jim Carrey, etc. However, it is important to note that Topics 8, 9, 11, and 12 center around the tech giants, such as Facebook, Google, and China's FaceSwap app Zao. These companies needed to act immediately to mitigate the risk posed by deepfakes. Topics 13, 16, 17, and 18 concern the improvements of the technology which needs assistance to identify the deepfakes and the creation of realistic deepfakes. 

Examining the comments in 2019, we found 18 topics generated through topic modeling. Topic 1 concentrated on the conversation of the deepfake pornography video ban on Reddit, the technology of creating it, and the legality of these actions. Unlike the posts, we found comments had similar patterns in discussion in Topics 2, 4, 6, 7, 8, 11, 12, 13, 16, 17, and 18. These discussions were more related to the deepfake video or images links shared in the posts and users were more involved with discussing the fact that generated deepfakes are indistinguishable from real. Topics 3 and 5 are discussions on news-related deepfakes and the technology behind deepfakes. Topics 9 and 10 regrad editing and creating deepfake videos, Topic 14 is involves moderating sub Reddits, identifying and detecting deepfake posts, and importantly, the actions that can be taken in regard to these posts. Topic 15 was about the conversations on how people believe deepfakes and the concerns this raises in society.

\subsubsection{Conversations during 2020}
In 2020, we found that deepfake related posts almost doubled compared to 2019. At the same time, the sub Reddits discussing deepfakes doubled as well. We found 19 topics from the corpus of 2020 posts. Topics 1 and 2 were posts about Donald Trump and Joe Biden's deepfake videos. Topics 3 and 17 focused on pornography, nudity, and related news where they gained greater interest than in 2019. Topics 4 and 7 were related to posts about Japanese deepfake memes and videos. This is the first emerging discussion on deepfake arising from Asia or any place other than the US. Topic 5 centered around deepfake related political misinformation and how Facebook was taking action against deepfake content. This was the first occurrence of the deepfake mitigation discussions by the platforms such as Facebook. Topic 6 was related to requests for deepfake videos. We could not find any interpretation for Topic 8 as it did not appearto be a meaningful discussion. Topics 9, 10, 11, 12, 13, and 18 from the posts had a similar pattern. These were about the deepfake video of Queen Elizabeth's Christmas message, Tom Cruise's deepfake videos, Elon Musk, KSI (a famous YouTube personality), and other famous personalities memes, videos, and music. On Topics 14, 15, and 16 we found posts relating to creating deepfakes, tutorials, and the types of tools that can be used to make realistic photos and videos. Finally, Topic 19 was specifically about the detection of AI technology for voice. 

Examining the comments of the 2020 Reddit corpus, we found 15 dominant topics. Compared to the number of comments in 2019, these did not double in growth. However, collectively the posts-to-comments ratio increased, which reflects the tendency to engage with deepfake posts. Similarly, as with the post, we found patterns in the 15 topics for comments. Topics 1, 3, 6, and 14 reflected discussions about how people were deceived by deepfake, how deepfakes are similar to realness, and how technology has improved to the level where no one can ever tell if it is deepfake without using technology. Topics 2 and 5 were related mostly to information requests from the users, and the moderation's conduct in Reddit.
Topics 4, 11, 12, and 13 were discussions based on detecting deepfake videos, which is an important discussion critically examining constructive evidence to the fake and the original. On Topics 7, 8, and 15, we could not interpret meaningful discussion. Unique to Topics 9 and 10, we interpreted the conversation as users being criticized for their deepfake video attempts, judging its degree of realism and users providing feedback to improve the content.

\subsubsection{Conversations during 2021}
Since we received posts and comments only from January 1 to August 21, 2021, we had fewer posts and comments compared to 2020. However, we found the trend to be increasing. Overall, discussions in 2021 were far more distributed in 1270 sub Reddits, whereas in 2020 only 1226 were found. Analyzing the topics generated by the post, we found ten dominant topics. Topics 1, 3, 5, and 9 had similar semantics. They collectively posted about nudity, sex, pornography, and arguably related websites where such images and videos could be purchased. We also saw India in the discussions of celebrity pornography. Compared to 2020, there was an increase of discussions and they were expanded to new geographic locations. Topics 2 and 6 were similar to 2020 viral videos of Trump and Tom Cruise deepfake related posts. Topics 4 and 10 were mostly related to creating deepfake using technology as well as building tools for detection of these deepfakes. Topic 7 was omitted as we could not trace a meaningful interpretation. 

Examining the comments during 2021, there was a slight reduction in the number of comments as our data were collected only until August of 2021. Adding four more months, we believe conversations about deepfakes were not diminishing but were growing with attention. The 2021 corpus provided 18 topics. However, we found that these discussions were spread with similar semantics. We categorize Topics 1 and 18 to be similar since both topics provide great insight into the possible occupations of deepfake. We believe this may have resulted due to the latest hiring of a YouTuber by a film production company~\cite{youtubehire}. Topics 2, 4, and 13 had similar discussions as previous years on regulating the sub Reddits, commenting on and removal of posts and comments. Topics 3, 6, and 14 depicted the concerns of deepfakes as a technology that can create distress among political figures, pornography, and technical concerns on how AI can be used to create these images and videos. Topics 7, 8, 9, 10, 12, and 17 were related to specific conversations based on the posts they were commenting on. This is a pattern that we came across in the 2020 comments as well, where users are provided critical reviews for their deepfake creations or the posts they submitted, not necessarily created by themselves. Similarly, Topics 5, 11, and 15 also provided insights on concerns in deepfake, and how it appears to be truly original. 

\subsection{RQ1-Second part: How have these conversations changed over time?}
\begin{figure}[t]
 \centering
 \includegraphics[width=\linewidth]{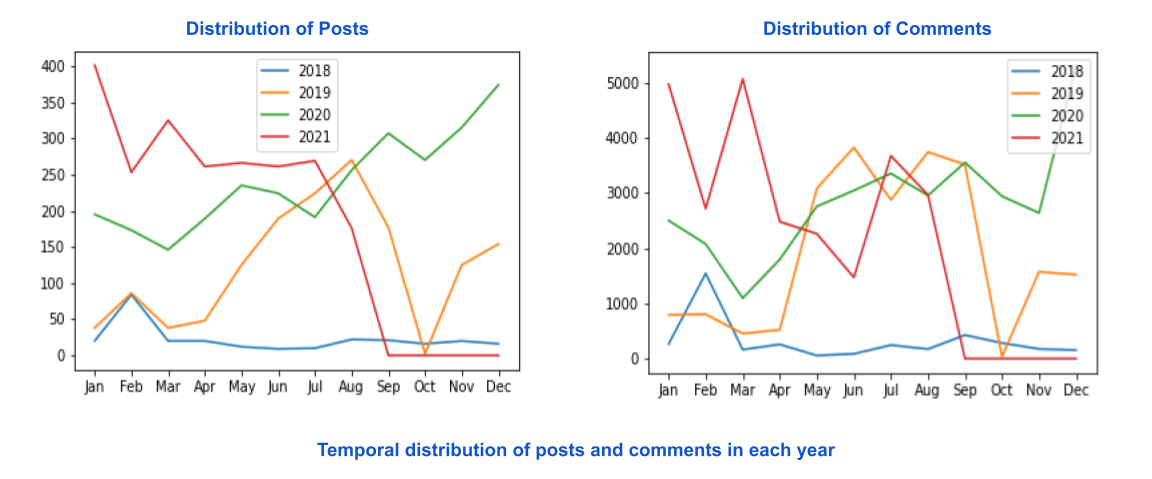}
 \caption{The temporal distribution of comments and posts over the years 2018, 2019, 2020, and 2021.}
\label{fig:temporal}
\end{figure}

We have previously stated the frequency of posts and comments when describing the topics throughout the years. To observe the frequency of post and comment behavior over the years, we projected a line graph (See Fig.~\ref{fig:temporal}). This graph shows the trends over each year. It is apparent that, 2021 had the highest number until early August, while 2018 had the lowest number each month. Table~\ref{tab:sum} summarizes the entire line of years with posts, comments, number of sub Reddits and the number of topics generated from each year, from posts and comments. There were 49 topics from posts and 66 topics from comments generated over the years. Examining and analyzing the key coherent words in each topic, we manually grouped them with similar patterns. For example, the topics related to moderation rules and regulations of Reddit became one group, and discussions about deepfakes concerns were grouped in another. This helped us to understand whether the discussions are the same each year and to find unique discussions among posts and comments over the years. As depicted in Fig.~\ref{fig:compare}, unique discussions are filled with color. 

We found unique posts from 2019 where users posted more on how platforms need to act to mitigate the misinformation from deepfakes. In 2020, unique posts were found about users discussing the creation of deepfakes. This went beyond the previous discussions where users not only discussed creating deepfake, but also its nuances, including formal tutorials on how to produce them. Users shared valuable resources and knowledge in building these videos and images with easily accessible tools and technologies. At the same time, 2020 uniquely brought Japan into the discussion based on certain memes created using deepfakes. This was mainly due to the hype brought by the ``Baka Mitai'' meme that features a Japanese song from the popular video ~\cite{memejapan}. In 2020, we also found unique discussions in posts focused on deepfake AI voice detection and constructive discussions on how to detect what are deepfakes and what are original videos. We believe that the discussions on AI voice detection may have arisen because, in 2020, an AI voice scam was determined to have manipulated the CEO of a company~\cite{aivoice}. Similarly in 2021, we found unique posts centered around pornography and celebrity deepfakes from an area of contributions, ``India''.  

Similar to the unique set of posts, we found unique reply behaviors in comments. In 2018, although there were not any unique posts, we found two unique discussions in comments---1) legislative actions against deepfake pornography and 2) self-cognition, the morality of deepfake. This was the first year that deepfake pornography videos were banned on Reddit. Subsequently, in 2019 we found discussions of news relating to deepfake technology. In 2020, the comments were more exclusively discussed on how to detect deepfakes due to the growth in creating them. In 2021, we found a unique set of comments about job hiring. We believe this was due to the news about a film production company hiring a YouTuber who had contributed realistic deepfake videos~\cite{jobhire}. 

Aside from the uniquely discussed posts and comments, we found continued discussions each year about concerns arising from deepfakes. These discussions depicted how platforms are failing to regulate and moderate deepfakes and how they might cause harmful consequences. However, apart from concerns, specifically during 2020 and 2021, we found that Reddit users were using the platform as a test-bed to review their created deepfake videos. This took place when they posted their deepfake video, and many other users provide critical feedback on how to improve the quality of the deepfake to reflect a more realistic effect. We also found the continued interest in discussions on how to create deepfakes by, usage of tools and methods where anyone can easily create them and specifically requesting deepfakes to be created for money or offering services to creating deepfakes for money. This led to Reddit space to become a marketplace for monetizing deepfakes.

\begin{table*}
 \caption{Summary of Posts, Comments, and topics distributed across the number of sub Reddits across the four years.}
\begin{tabular}{r|r|r|r|r}
\hline
 Year & Number of Posts & Number of Comments & Number of sub Reddits & Number of Topics in posts and comments \\
 \hline
 2018 & 269 & 3828 & 165  & 1, 15    \\ 
 2019 & 1472 & 22773 & 538 & 19, 18   \\ 
 2020 & 2886 & 34071 & 1226 & 19, 15  \\ 
 2021 & 2280 & 25651 & 1270 & 10, 18   \\ 
 \hline
 \end{tabular}
 \label{tab:sum}
\end{table*}

\subsection{RQ2: What are the societal implications of these conversations?}

Qualitative open codes were used to derive possible societal implications from these conversations.  An open code represents one or two sentences describing the actions of a topic resulting from topic modeling. The coders first observed each topic and the obtained samples related to the topic. These samples provided a better sense of the discussion topic and how such discussions may lead to implications that may not be directly visible through topics. Then they wrote a short description of possible outcomes discussion on a topic. For example, based on the posts in 2019, an open code for Topic 1 indicated a sentence such as ``Creating original-like deepfake music video to a friend''. This was based on the keywords in Topic 1 (video, music, show, original, real, full..etc) and the related Reddit post samples obtained from Topic 1. We continued this process from each topic in each year collectively for all Reddit posts and comments. We conceptualized each topic to an action where there was a possible reaction attached to it. All together we had 135 topics---49 topics generated from posts and 66 topics generated from comments that coders independently observed to generate open codes. Each coder came up with 59 and 73 open codes which described possible implications respectively.  
After both coders observe and reviewed all topics and sample conversations, all researchers convened to interrogate and debate on the open codes. Based on the similarities of the outcomes and actions, we categorized these codes into ten themes. Explaining the detailed categorizations and the process of deriving the implications are beyond the scope of this paper. However, we next, explained each from a macroscopic viewpoint as it reflects a possible implication from these conversations. Details of qualitative coding and categorizing can be found in our open science framework link---\url{https://osf.io/dg2wh/}.

\changes{\textit{1. Reddit space becomes an on demand archive for banned deepfake videos}---On occasion, social media platforms ban posts relating to deepfake videos or images due to violation of platform rules. Some users often miss this banned content and seek help in the sub Reddits to find support in locating these content. Deepfake communities in Reddit often reserve their spaces for such inquires and share with each other when someone needs to revisit the content. Although these videos may have been banned from the platforms, Reddit communities help to retrieve them in any form. In some cases, we observed that users saved the banned content in the local machines and shared them via emails when requested. In other cases they were provided the information on how to re-create such banned content in their machines. Therefore, the community in Reddit serve as an archival for deepfakes.}

\changes{\textit{2. Reddit space incubating deepfake creators}---Reddit community space for deepfakes provides an incubator for developing deepfake generators/creators. The community often provides critical feedback on how to improve the models, which tools to use, or often shares tutorials and knowledge on how to do it effectively. There were significant occasions that users (possibly deepfake creators themselves) often posted created deepfakes and received feedback for quality. They often receive useful tips to create better quality deepfakes either using new technologies or tools. These requests and replies create a safer place for learning and optimizing the knowledge for creating deepfakes.}

\changes{\textit{3. Reddit becomes a collective space to question and raise concerns about deepfakes}---Primarily, Reddit has depicted a pro-deepfake space and the  majority of users were specifically creating deepfakes and majority praised the technology in this space. Although Reddit provides a very unique supportive space to creating deepfakes, there are numerous occasions that the users discuss the consequences of deepfake productions, and raise their concerns. Discussions revealed that each year, some have utilized the Reddit as a platform to raise their concerns on the technology in general, specifically concerns about the deepfake content created and shared on the internet. The magnitude of the concerns raised in this community could involve changes in Reddit policies and subsequently to other social media platforms.}

 \changes{\textit{4. Reddit expert users support anyone to create customized pornography}---Deepfake pornography is banned on many platforms. However, it is banned only if it is published as pornography. Experts in Reddit communities share all the knowledge, tools and components (such as databases to purchase images) to create porn related deepfakes from scratch. The Reddit space is continuously used as a test-bed for any novice to experiment while creating deepfake videos. It also provides ample inspiration for deepfake sources and target combinations, such as the recent introduction of deepfake Indian celebrity pornography, or Japanese cartoon characters. In other words, the platform provides the recipe of how to create your own deepfakes more easily, specifically pornographic videos.}

 \changes{\textit{5. Reddit users provide inspirations for deepfake movies and knowledge on surrounding legal clearances}---Deepfake conversations related to movies have mostly been in regard to opportunities for the  entertainment industry from deepfakes. Most opportunities suggests having films with or without real actors, and voices, or new characters artificially created for movies using deepfakes. As a sidebar to the opportunities, conversations also focused on the consent and legality of future AI-created actors. This provides users new knowledge and a network to discuss such unexplored and undefined areas in the entertainment industry.} 
 
 \changes{\textit{6. Reddit users ingenuity to mix variety of sources with a variety of targets in creating deepfakes}---The technology of deepfake provides ample opportunities to mix sources and targets with a wide variety of samples. In the real world, mixing the DNA of real humans is prohibited and costly. However, with deepfake, mixing different entities provides opportunities for entertainment. Since these can be created easily with the many tools available as discussed in Reddit communities and as they can be tested easily by reviewers and feedback can be obtained rapidly, Reddit space is creating an emerging  market for it. One such example is the deepfake clip using the Hollywood film star Keanu Reevese in a Bollywood movie~\cite{KanueBolliwood}. Reddit communities provided faster feedback and many inspirations on how to make unusual possibilities.}

 \changes{\textit{7. Reddit platform and other social media platforms disproportionately control their deepfake policies}---Reddit banned deepfake pornography in 2018, while Facebook banned deepfakes in 2020. At the same time, each platform regulates the content in their own way, and removing and banning synthetic content has brought controversy concerning the actions within platforms. Such inconsistent time frames and policies provide many opportunities to do enough damage. For example, by the time Facebook banned deepfakes, it was only for pornography, but other sorts of harmful actions, such as creating deepfakes to threaten personal disputes, and harm democracy have not been properly identified. }
 
 \changes{\textit{8. Reddit community is more concerned with deepfakes videos but less concerned with the deceiving power of AI-generated text}---Reddit communities feature extensive conversations about the concerns arising from deepfakes and one of the major emphasis carried on involves AI-generated news. It is worth to mention that, although AI helps people detect misinformation, it has ironically also been used to produce misinformation. Transformers, like BERT, and GPT use NLP methods to understand the text and produce translations, summaries, and interpretations. The capabilities of the technology have also been extended to generate artificial news with similar legitimacy (e.g, www.NotRealNews.net). Although, the community in Reddit highlight the platforms limited capabilities to identify any AI-generated content, we found they have not paid sufficient attention to sever issues such as AI-generated news or artificially created content in platforms. However, they have paid significant attention to AI-generated videos and concerns over it.  However, the community has ignored the harms that may occur from AI-generated content that could be used for fake reviews, or personalizing fake emails forging identity etc,. }

 \changes{\textit{9. Reddit and other social media platforms influence elections with deepfakes}---Reddit had many discussions based on the deepfake videos created involving key political figures such as Obama, Trump, and Putin. On top of the fact that videos being deepfake, the amount of likes and trends they get on the platforms brought concerns into the discussions. These viral videos may impact the elections as the public has often been deceived by deepfakes as discussions have explained in Reddit.  As researchers suggest, since the majority of platforms are incentive-driven based on engagements such as likes, views and shares, we witness much content purely created for entertainment, yet using political leaders who mostly made controversial news. This trend may likely create an impact on the social behaviors of users and impact the choice of decisions they make in everyday life.}

 \changes{ \textit{10. Reddit users create a deepfake marketplace}---Deepfake related job hiring has been a long-discussed topic on Reddit.  A popular deepfake YouTuber who goes by the name ``Shamook'' has been hired by Lucasfilm corporation. Shamook's most viral video is a deepfake that improves the VFX used in ``The Mandalorian'' Season 2 finale to de-age Mark Hamill’s Luke Skywalker. The company says they have been investing in both machine learning and AI as a means to produce compelling visual effects work and they found it to be terrific to see momentum building in this space as the technology advances~\cite{youtubehire}. Similarly, there were many other discussions related to new jobs such as deepfake video creation as a service, selling deepfake related videos to a value and to some extent it was visible that some websites were selling images to be created as deepfake. For example, we found posts requesting to create customized deepfakes for money and the replies contained emails and at some point shared their previous creations as proof of the quality of service. Not only selling as a service, Reddit conversations included websites that offer deepfakes as a service, where users can register to consume deepfakes, mostly pornographic videos. The pro-deepfake culture within Reddit communities and the nature of specific tech user groups and the exchanges of deepfake have created some sub Reddits as a marketplace for deepfakes where anyone can request anything related to deepfakes to be created for money.}

\section{Discussion}
Our findings revealed a variety of emerging topics about deepfakes in Reddit communities. Unlike other platforms, Reddit serves a niche community~\cite{may2019ask} that is generally interested in deepfake and the members of the community. This community is anonymous to each other. The discussions discovered on Reddit may not be fully visible on other social media platforms, such as Twitter or Facebook, because these communities are not anonymous and thus users tend to avoid discussing intimate details. We observed an incremental distribution of posts and comments on deepfake each year. Although we did not concentrate on particular sub Reddits but rather the general conversations from the platform, we found that the post-to-comment ratio was at a higher level leaving an effective engagement pattern in many sub Reddits. This reflects the interest of Reddit users to discuss deepfake related topics. Although we can not generalize the main findings obtained from Reddit to the wider world, our large-scale data sample represents a considerable amount to foresees the implications that may arise from conversations of deepfake in the future. The directions of the discussions may lead to unprecedented societal harm unless platforms and technology designers take preventive actions ahead of time.

\subsection{Are deepfakes concerning?}

Analyzing the conversations about deepfakes over the years, we found topics which may lead to concerning situations over its consequences. Specifically, when there was Reddit ban on pornography in 2018, we found unique discussions driven towards questioning the morality of creating deepfakes. This concern was seen in 2019. However, it was not visible in as a key topic in 2020 but appeared again in 2021. This reflects that the community is gradually normalizing the deepfake artifacts and does not foresee the harm; rather there is excitement about creating more. On the other hand, an important finding from these discussions centering around ``concerns'' was that many users in Reddit depicted their fears towards ``political distress'', ``election propaganda manipulation'' and ``pornography''. As research revealed these are well anticipated social impacts~\cite{brooks2021popular}. We could not find strong discussion topics about the risk to individuals, personal abusel, or reputation decimation as found in recent research that analyzed popular news articles about deepfakes~\cite{brooks2021popular}. However, we believe such discussions, i.e.,---personal attacks using deepfakes, reputation damages and financial attacks may not be visible in Reddit conversations, since such conversations may appear in personal conversations or private groups. At the same time, the field is new and only a niche community knows the nuances of using it for personal benefit. In our analysis, throughout the years from 2018 to 2021, we found strong moderation on deepfakes and this could also be another another reason that we did not encounter such discussions in the community.

Apart from users fearing the improvements of this technology in 2018 and 2020, we observed that Reddit communities explored the potential of deepfakes in their discussions---pointing out how to create and edit deepfakes using tools and technology trending as early as 2018. This trend has even led into unique conversations where they provide tutorials on how to create deepfakes. We imply that Reddit can be viewed as an incubator for nourishing deepfake creators, which poses a series of concerns, i.e.,---Reddit communities shared Udemy tutorials \url {(https://www.udemy.com/course/deepfakes/)} on creating deepfakes with practical hands-on applications shared by users. Although the instructor warns the participants against doing harm, no preventive mechanisms are taken or mentioned in the Reddit communities. Each year we found trending topics where users discussed Reddit bans, moderation of deepfakes, and bots removing some posts yet it appears that users enjoy these discussions, specifically critics of the created posts and context of the deepfake. This enables a positive environment for potential deepfake but less concern about the impact it can create. 

At the same time, we see discussions on how the platform needs to take action against deepfakes, yet policies seem to be mostly limited to curbing pornography~\cite{maddocks2020deepfake}. The areas of security and politics and entertainment are under severe threats which need platform policies and legal frameworks. Reddit discussions turn to such topics occasionally, but since the platforms are incentive driven for trends, likes and the sharing culture, we see a continuing interest to creating trending videos using deepfakes.

Through our topical explorations, we found deepfake discussions leading to a marketplace that monetizes deepfakes. This includes conversations offering to create customized deepfakes for money, requests to get done certain customized deepfakes and sharing websites where users can consume deepfake products for money such as pornography. Although monetizing deepfake pornography has been around since 2018 ~\cite{wagner2019word}, our research found new forms of a marketplace where users showed their interest in creating deepfakes as a service. These services include custom-made deepfake with a specific purpose such as entertainment clips, political figures talking or personal documentaries. Users publicly discussed pricing and exchanged their information in discussion. At the same time, there were many conversations centered around entertainment, specifically preparing to invest in film industry using AI presence or voice of certain actors and actresses. It is no longer secret that there is a wide range of deepfake service platforms, i,e.,---websites that provide a user interface  to help users accelerate the process of creating deepfakes. These websites typically require users to upload training data in the form of media objects pertaining to the subjects they intend to deepfake. They also provide tools for receiving the AI-generated deepfake media object once it has been created. These service platforms may handle all of these back-end functions in an entirely automated fashion, or through using a partially manual processing by the service’s employees or contractors. These websites functioning for a fee in providing such deepfake services and such information has been shared in Reddit communities. 

Our derived implications lead towards more concerting directions. However, we must emphasize that deepfake technology is a double-edged sword. It has many positive uses to the society and must not work towards eliminating them rather mitigating the risks that come with the technology. That will be a fiendishly difficult balance for society to make, especially as deepfake tools evolve to eliminate any trace of how they have altered videos, images, audio, and other media at the center of our lives in the 21st century. Therefore, clearly, it is time that AI-generated fake content will deeply impact our lives. In all sense, we can consider deepfakes to be a ``dual-use'' technology, just as likely to be used for good as for evil. As researchers pointed out \textit{``If fake videos are framed as a technical problem, solutions will likely involve new systems. If fake videos are framed as an ethical problem, solutions needed will be legal or behavioral ones"}~\cite{brooks2021popular}. We invite researchers to view this greater societal problem and encompass solutions using a multi-disciplinary viewpoint.

\subsection {Implications for Human Computer Interactions}

\subsubsection {Theoretical implications} 
Our work has presented an empirical understanding of the Reddit ecosystem that endorse deepfakes regardless of its positive or negative implications. We believe this is the first large-scale analysis of deepfakes conversations in online communities. Our findings shed light on what Reddit communities discuss about deepfakes and, how might these conversations impact society at large. However, the possible implications centered upon deepfakes conversations on Reddit provide a challenging situation to take preventive actions since some issues have not matured to indicate severe harm to the society yet. For example, since Reddit communities are sharing previously banned deepfakes, they act as an archival space. This may not immediately cause harm depending on the obtained artefacts. In addition, Reddit communities acting as an incubator for deepfake creators, encourages many aspiring tech savvy people to create more and more deepfakes regardless their harmful side effects.  

One major topic revealed that Reddit has well-established moderation strategies, including rules and guidelines, and bots to regulate sub Reddits and bad behaviors in online communities~\cite{fiesler2018reddit}. However, we observed some of the communications in Reddit may not be against rules or bad behavior but subject to moral values. Example includes a user requesting a deepfake video which already been banned from the platforms; a user offering to create deepfake content for money; and a user offering intellectual help to create deepfakes without knowing whether the knowledge obtained will create harmful content. Future work could examine this apparent but untouched area of capturing the morality of the conversations, either by educating the moderators and users of potential harm and building interventions for users before they post or reply to items in specific category. Capturing morality is a challenge to many NLP scientists and this needs interdisciplinary research collaborated with HCI researchers to understand the nuances in such communities. 

Currently much research has focused on automated approaches for the moderation of online communities, yet there are not sufficient studies to understand certain behaviors of communities which may cause harmful consequences. There is also a lack of studies on the enforcement side, and considerations to be made during this process. Enforcement strategies depend on the platform and the specific community where strategies vary significantly in intent and execution~\cite{chandrasekharan2018internet}, and these nuances should inform detection strategies in the future. Our work provides the nuances of certain conversations that cause implications to society. Such topical exploration inquires in platforms to understand how to take the next step.  New AI-based moderation system can be easily customized to detect potential harmful content that violates a target community’s norms and enforces a range of moderation or design informative actions.

\subsubsection{Design implications}
Our study sheds light on designing HCI systems and processes to mitigate the harmful nature of deepfakes. Specifically, this includes discussions of topics about platform moderation, tools, and bots regulating deepfakes, and people's concerns about deepfakes of nudity, pornography, and misinformation which reflect the negative perspective of deepfakes. However, creating deepfakes, accessing archival deepfakes, using the community as a test-bed to obtain feedback, helping to learn about creating deepfakes, and sharing deepfakes and deepfake related jobs reflect the positive perspective of deepfakes by the community yet implications may lead to negative outcomes. Considering the possible implications of these discussions, our study provides design implications that HCI researchers could work on to mitigate the harm created by deepfakes.

Social media platforms need more advanced policy designs to mitigate damages cause by deepfakes. Community users in Reddit have continually echoed this proposal and not just by policies but also design decisions to mitigate deepfakes in platforms. For example, platforms have taken numerous actions against misinformation such as nudging in the user interfaces design~\cite{konstantinou2019combating} (designing subtle interventions to guide the user with choices without restricting them). However, deepfakes should be considered as a new domain and users need to be informed before they take any actions. 

Our findings suggest that some conversations may lead to harmful implications, which may not be detected by bots or human moderators. It is essential to establish mechanisms to report potentially harmful activities on deepfakes, as well as legal frameworks to support actions against deepfakes designed and linked through platforms.


Our research determinded that Reddit users extensively discuss creating deepfakes and getting help, thus, incubating space for deepfake creators to produce quality deepfakes. However, it is equally important to provide education and knowledge on deepfakes, their possible social harms and detecting these videos. One effective example was presented in CHI2021 where authors demonstrated a carefully crafted training---based on how the human eye deceive deepfake~\cite{tahir2021seeing}. This was the first attempt of HCI researchers to tackle deepfakes from the users point of view. However, we seek more solutions integrated to the platforms when such topics are discussed such as prompting nudges, educating on consequences and providing psychological readiness for defeating the negative impacts of deepfakes. Since the users are more into creating deepfakes where commercial tools are at easy access, we highlight the need of regulating AI tools that provide such access. The researchers should seek for novel methods to distinguish real vs fake. One way is to look at embedding precautions to the tools and for users. Watermark method "Markpainting" is one the novel method that researchers believe can be used as a manipulation alarm that becomes visible in the event of inpainting~\cite{khachaturov2021markpainting} and can be integrate to detect any manipulations in platforms. Similarly, HCI and policy researchers can aid in designing deepfake detection's integrated into social media platforms and regulatory measurements on the shared deepfakes platforms to warn the users. This is vital considering the growth in usage of AI tools for deepfakes.
Another possible intervention that HCI researchers should consider is to regulate deep learning tools and research under strict ethical guidelines. The majority of AI tools are implications of the advancement of AI and deep learning technology research. Such research poses major societal risks, yet many of the institutions' Ethical Review Boards (ERB) or similar entities do not require any prior approvals to conduct AI research, because such research usually does not focus on human subjects directly. However, we emphasize the need for designing regulatory measurements for such research and industry applications imposing preventive actions. One such inspiration has been prototype by introducing Ethics Society Review (ESR) board for ML/AI projects~\cite{bernstein2021esr}. We stress the need to popularize this review mechanism in both industry and academia for the betterment of society.

\subsection{Limitations of this research}

Our approaches hinges on the NLP process and the algorithms we used in topic modelling. Our key results---i.e., deriving topics based on conversations and then using these topics to derive implications may contain bias on algorithmic decisions. The algorithm provided are the set of prioritized words based on similarity for a set of topics. Although manually it is not possible to sort a large sample of text and categorise it into topics, when identifying concepts from a collective of topics, many have to rely on the certain weight distributed in words. Our algorithm may have concealed emerging important discussions which may contribute into topical discussions, yet due to less probabilistic nature it may be not possible to pick on thresholds. 

Our qualitative codes were based on two researchers and we did not concentrate on inter rater reliability, but collectively we expanded and reduced some codes based on collective agreement. Our results are solely based on deepfake conversations on Reddit, hence the implications and proposed mitigation actions are narrowed from what we observe during these discussions, where there could be prominent implications that may not appear in these discussions, therefore there can be a myriad of other possible future directions.

\section{Conclusions}
In this study, we explored the deepfake conversations on Reddit over the last four years using a mixed-method approach. We first used a topic modelling in NLP to understand the conversation topics, and then analyzed and compare topics in each year from 2018 to 2021. We then used qualitative coding to map implications from the conversations. Based on overall conversations out of all four years, we found pornography, political discourse, concerns on ``fakeness'' and conversations on Reddit moderation were dominant. However, analyzing each year's conversations, we found some topics are unique to each year (i.e., self-cognition and questioning the morality of deepfakes in 2018, constrictive discussions on how to detect deepfake in 2020, etc). Other conversations have been continued for years (i.e., concerns arise from deepfakes, conversations about deepfakes been moderated by bots and moderators). We derived 10 implications based on the conversations and our conclusion remarks included that conversations of deepfakes have provided opportunities and a concerning  future. Finally, we provide implications where deepfakes societal harm caused by deepfakes can be mitigated in the long run. However, we stress that mitigating deepfake concerns will not be a single occurance and achieving this goal requires multidisciplinary collaboration.

\begin{acks}
We would like to thank Professor Michael Bernstein at HCI group Stanford University for the feedback and support provided to address reviewer comments and to articulate the societal implications of deepfakes from this study. This work was generously supported by JST, CREST Grant Number JPMJCR20D3, Japan.
\end{acks}

\bibliographystyle{ACM-Reference-Format}
\bibliography{samplebase}


\appendix
\section{Appendix}
\subsection{Example topic coherence}

\begin{figure*}[]
 \centering
 \includegraphics[width=\linewidth]{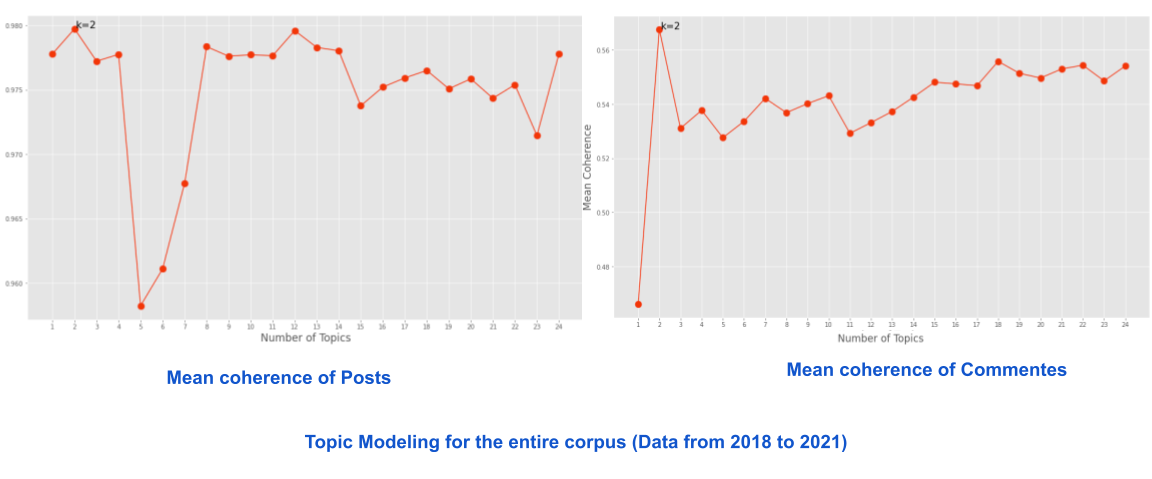}
 \caption{}
\label{fig:coherence1}
\end{figure*}

Figure~\ref{fig:coherence1} shows an example of using NMF Topic coherence $k$ which indicates the optimal number of topics. In this case,there are four dominant topics visible in the entire corpus---2 from posts and 2 from comments. 

The NMF Topic modeling resulted two dominant topics from each corpus of "Posts" and "Comments" in all years together (2018 to 2021) (shown in Table~\ref{tab:coherence2}. The first 2 topics are generated from entire posts and a second set of two topics generated from comments of all corpus.

\begin{table}[]
    \centering
    \begin{tabular}{lrp{5.7cm}}
        Source & \# & Terms \\
        \hline
        Posts & 01 & video, trump, someone, know, original, face, music, donald, create look\\
         & 02 & porn, videos, nude, fake, sex, tech, ban, face, ai, celebrity\\
        Comments & 01 & fake, people, think, deep, look, get, see, vi
        deo, know, deepfake\\
         & 02 & https, link, post, please, bot, subreddit, question, moderators, automatically, contact 
    \end{tabular}
    \caption{}
    \label{tab:coherence2}
\end{table}


\subsection{Detailed topics}
\begin{figure*}[]
 \centering
 \includegraphics[width=\linewidth]{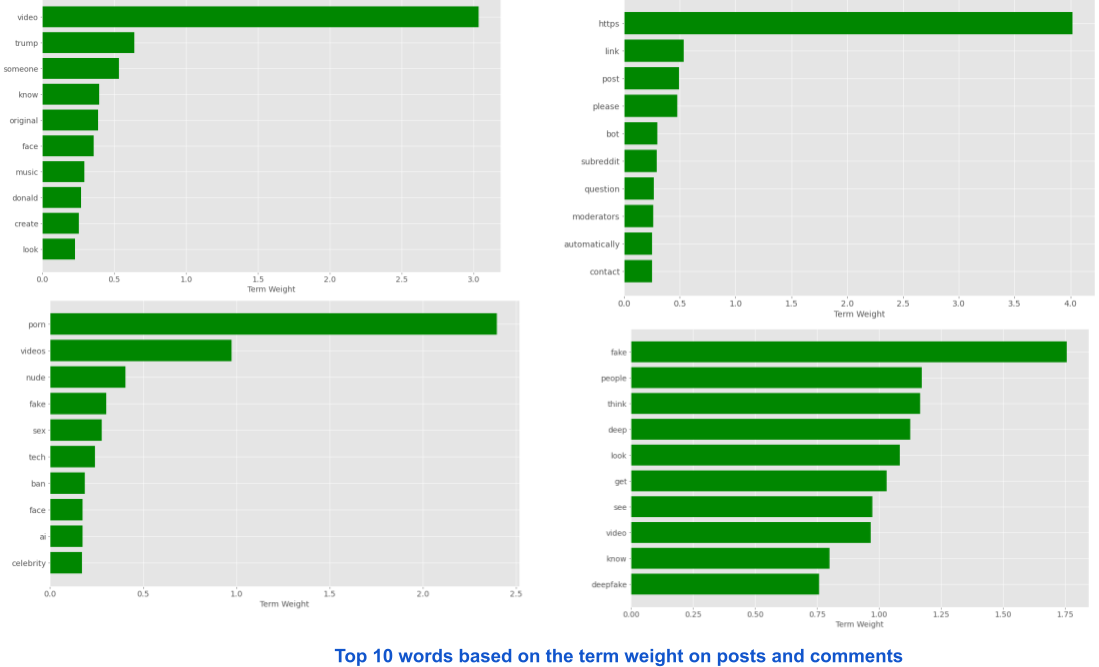}
 \caption{}
\label{fig:coherence3}
\end{figure*}

Each topic had certain words with weights based on the coherency to the topic. For example figure~\ref{fig:coherence3} reflect the top 10 key coherent words from the entire corpus.

The detailed semantically coherent words for Topic 1 generated from posts in 2018 using NMF model are shown in Table~\ref{tab:conversation20181}. 

\begin{table}[]
    \centering
    \begin{tabular}{rp{6.7cm}}
        \# & Terms \\
        \hline
        01 & deepfake, videos, video, someone, porn, cage, technology, https, news, tech \\
    \end{tabular}
    \caption{}
    \label{tab:conversation20181}
\end{table}

The detailed semantically coherent words for 17 Topics were generated from comments in 2018 using NMF mode are shown in Table~\ref{tab:conversation20182}.

\begin{table}[]
    \centering
    \begin{tabular}{rp{6.7cm}}
        \# & Terms \\
        \hline
        01 & think, something, need, really, immoral, moral, thing, right, person, things \\
        02 & https, http, bot, deepfake, time, train, comment, sure, example, version \\
        03 & look, kinda, hila, ethan, guy, scene, alike, tho, put, orton \\
        04 & get, better, stuff, news, want, way, attention, work, point, need \\
        05 & face, body, put, replace, swap, person, someone, image, thats, right \\
        06 & fake, deep, real, videos, news, image, tell, detect, convince, detection \\
        07 & video, evidence, footage, technology, edit, go, work, take, camera, court \\
        08 & please, post, question, automatically, link, subreddit, contact, http, moderators, remove \\
        09 & people, believe, want, real, problem, things, videos, take, truth, trust \\
        10 & de, que, un, na, la, pas, en, le, les, uma \\
        11 & good, really, thing, pretty, point, tell, go, probably, need, oh \\
        12 & reddit, ban, sub, go, rule, media, admins, mods, time, yeah \\
        13 & porn, deepfake, someone, lot, revenge, legal, celebrity, real, consent, celeb \\
        14 & see, fuck, want, things, shit, go, black, man, something. mirror \\
        15 & know, deepfakes, happen, man, future, lol, show, oh, kinda, shit \\
    \end{tabular}
    \caption{}
    \label{tab:conversation20182}
\end{table}




The detailed semantically coherent words for 19 Topics generated from posts in 2019 using NMF model are shown in Table~\ref{tab:conversation2019}.

\begin{table*}[]
    \centering
    \begin{tabular}{rl}
        \# & Terms \\
        \hline
        01 & video, music, show, original, real, full, look, something, disturb, cnet \\
        02 & videos, threat, detect, real, experts, world, warn, race, fear, election \\
        03 & porn, revenge, sex, world, virginia, politics, anal, celebrity, online, best \\
        04 & trump, donald, bean, president, evil, call, audio, boris, publish, putin \\
        05 & tom, cruise, bill, hader, morph, channel, man, iron, show, downey \\
        06 & face, swap, shift, ctrl, future, somone, actor, elon, people, stallone \\
        07 & jim, carrey, shining, jack, star, episode, asian, scary, psycho, dwight \\
        08 & technology, get, movie, good, able, david, future, actors, realistic, work \\
        09 & zuckerberg, facebook, mark, instagram, policy, take, ceo,' challenge,' sinister, reuters \\
        10 & joe, rogan, theo, von, brendan, chris, schaub, cage, nicolas, bert \\
        11 & deepfakes, google, ai, fight, show, vs, spot, work, detect, rule \\
        12 & app, zao, chinese, viral, privacy, go, china, concern, dicaprio, leonardo \\
        13 & voice, ceo, actors, podcast, talk, poetry, transfer, peterson, president, actor \\
        14 & keanu, reeves, look, real, bollywood, vfx, corridor, scene, creating, bob \\
        15 & fake, ai, news, deep, life, obama, sinale, everyone, person, barack \\
        16 & see, best, get, think, longer, time, go, inside, believe, guy \\
        17 & someone, need, help, know, get, please, real, find, creator, original \\
        18 & create, https, reuters, artists, techreview, software, realistic, rt, take, reality \\
        19 & tech, election, scary, social, company, future, media, bbc, house, google
    \end{tabular}
    \caption{}
    \label{tab:conversation2019}
\end{table*}


The detailed semantically coherent words for 18 Topics generated from comments in 2019 using NMF model are shown in Table~\ref{tab:conversation20192}.

\begin{table*}[]
    \centering
    \begin{tabular}{rl}
        \# & Terms \\
        \hline
        01 & deepfakes, porn, technology, real, videos, detect, ban, create, start, illegal \\
        02 & https, link, fresh, expire, discord, invite, work, check, join, youtube \\
        03 & fake, deep, videos, news, real, voice, tell, claim, detect, trump \\
        04 & look, real, better, arnold, bad, hader, eye, bill, head, lol, better, audio, technology, porn \\
        06 & think, mean, yeah, point, way, person, thing, lol, better, real\\
        07 & get, better, ta, try, ban, little, worse, rick, back roll\\
        08 & see, want, best, movie, believe, Ya, try, ban, little, worse love, wait, pretty, far \\
        09 & video, evidence, edit, real, trust, watch, audio, court, trump, source \\
        10 & face, voice, swap, put, body, replace, change, bill, different, hader \\
        11 & good, pretty, thing, job, sure, ones, damn, yeah, imagine, idea \\
        12 & know, thing, want, feel, guy, everyone, song, watch, probably, comment \\
        13 & go, way, happen, back, technology, future, lot, thing, live, far \\
        14 & post, please, question, subreddit, comment, automatically, moderators, bot, action, contact \\
        15 & people, believe, videos, want, things, bad, real, problem, care, world \\
        16 & fuck, shit, na, gon, holy, watch, yeah, give, man, oh \\
        17 & time, need, work, something, take, ai, right, someone, image, want \\
        18 & really, cool, wow, great, feel, want, amaze, interest, weird, watch \\
    \end{tabular}
    \caption{}
    \label{tab:conversation20192}
\end{table*}


The detailed semantically coherent words for 19 Topics generated from post in 2020 using NMF model are shown in Table~\ref{tab:conversation2020}.

\begin{table*}[]
    \centering
    \begin{tabular}{rl}
        \# & Terms \\
        \hline
        01 & video, music, original, create, biden, funny, release, post, show, tweet \\
        02 & trump, donald, biden, office, president, speech, joe, lack, park, parody \\
        03 & porn, internet, celebrity, bella, lack, think, revenge, find, sex, bad \\
        04 & mitai, baka, help, kiryu, yakuza, choir, original, rick, jack, memes \\
        05 & videos, facebook, ban, election, tech, https, remove, go, ahead, misinformation \\
        06 & someone, please, need, request, pls, get, look, reddit, tell, help \\
        07 & dame, da, ne, bakamitai, yakuza, adam, dane, learn, really, link \\
        08 & pewdiepie, wolf, street, wall, ludwig, magicofbarca, felix, pewds, jacksepticeye, credit \\
        09 & queen, christmas, message, channel, elizabeth, alternative, deliver, speech, ii, warn \\
        10 & tom, future, holland, back, robert, downey, cruise, jr, rdj, star \\
        11 & know, original, want, jj, girls, kik, good, look, get, source \\
        12 & sing, song, bakamitai, musk, elon, choir, star, give, hate, stalin  \\
        13 & ksi, pokimane, ainsley, jj, meets, magicofbarca, scary, please, movies, post  \\
        14 & face, need, look, swap, app, put, try, guy, old, tutorial  \\
        15 & deepfakes, create, post, discussion, good, voice, app, want, tell, adversarial  \\
        16 & best, see, get, go, way, want, probably, far, real, photo  \\
        17 & fake, deep, real, kik, news, image, nudes, girls, create, people  \\
        18 & meme, request, funny, time, go, big, rush, help, take, king  \\
        19 & ai, technoloav, music, detection, voice, request, world, musk, elon, think
    \end{tabular}
    \caption{}
    \label{tab:conversation2020}
\end{table*}


The detailed semantically coherent words for 15 topics generated from comments in 2020 using NMF model are shown in Table~\ref{tab:conversation20202}.

\begin{table*}[]
    \centering
    \begin{tabular}{rl}
        \# & Terms \\
        \hline
        01 & people, believe, real, deepfakes, news, things, media, problem, lie, want \\
        02 & https, link, download, youtube, info, message, find, bot, help, google \\
        03 & fake, deep, real, videos, news, technology, tell, ai, detect, porn \\
        04 & look, better, eye, real, original, cgi, weird, great, realistic, tom \\
        05 & please, post, subreddit, automatically, question, moderators, bot, contact, action, perform \\
        06 & deepfake, porn, technology, better, deepfakes, someone, original, videos, real, ai \\
        07 & see, best, love, want, movie, wait, thing, na, believe, comment \\
        08 & think, mean, lol, fit, something, guy, part, honestly, saw, thing \\
        09 & good, pretty, job, bad, deepfakes, damn, idea, thing, quite, point \\
        10 & get, better, fuck, shit, na, gon, ta, try, lol, point \\
        11 & video, evidence, real, original, audio, watch, link, youtube, videos, deepfakes \\
        12 & know, fuck, thing, mean, real, happen, exist, right, oh, love \\
        13 & face, voice, put, body, head, model, swap, actors, match, actor \\
        14 & really, great, cool, impressive, wow, interest, watch, nice, love, movie \\
        15 & go, need, time, work, take, way, something, right, want, someone
    \end{tabular}
    \caption{}
    \label{tab:conversation20202}
\end{table*}


The detailed semantically coherent words for 10 Topics generated from posts in 2021 using NMF model are shown in Table~\ref{tab:conversation2021}.

\begin{table*}[]
    \centering
    \begin{tabular}{rl}
        \# & Terms \\
        \hline
        01 & porn, sex, fake, sexy, hot, actress, celebrity, indian, bbc, videos \\
        02 & video, trump, original, music, find, sex, name, face, evidence, latest \\
        03 & nude, sex, actress, edit, south, sell, paypal, ha, selling, ilpt \\
        04 & someone, look, please, face, looking, create, find, movie, really, help \\
        05 & nudes, free, bot, best, create, https, good, site, website, want \\
        06 & tom, cruise, videos, tiktok, real, viral, realistic, create, tech, go \\
        07 & en, para, los, link, comments, fuck, de, una, comment, dando \\
        08 & ai, voice, need, sing, technology, think, eminem, detection, sample, sound \\
        09 & know, want, kik, dm, girl, pics, original, girls, need, picture \\
        10 & get, deepfakes, technology, see, post, ban, look, best, detection, fuck
    \end{tabular}
    \caption{}
    \label{tab:conversation2021}
\end{table*}


The detailed semantically coherent words for 18 Topics generated from comments in 2021 using NMF model are shown in Table~\ref{tab:conversation20212}.

\begin{table*}[]
    \centering
    \begin{tabular}{rl}
        \# & Terms \\
        \hline
        01 & people, want, real, deepfakes, believe, porn, need, mean, lot, things \\
        02 & https, check, link, channel, article, want, source, github, free, discord \\
        03 & fake, deep, real, tell, trump, news, detect, videos, porn, technology \\
        04 & please, subreddit, automatically, question, moderators, bot, contact, concern, action, perform \\
        05 & look, real, exactly, luke, bad, weird, kinda, hair, kind, something \\
        06 & deepfake, porn, someone, technology, tech, videos, image, need, software, de \\
        07 & think, thing, something, guy, saw, yeah, mean, right, sure, way \\
        08 & see, best, love, want, thing, internet, wait, sub, something, great \\
        09 & know, yeah, mean, tell, guy, talk, everyone, someone, old, right \\
        10 & good, pretty, job, thing, yeah, man, point, bad, love, damn \\
        11 & video, original, watch, real, edit, trump, youtube, link, game, frame \\
        12 & get, ta, guy, try, probably, help, away, ban, little, old \\
        13 & post, comment, remove, sub, Link, read, rule, twitter, thank, reddit \\
        14 & face, voice, body, put, actor, someone, head, ai, original, actors \\
        15 & really, cool, great, hope, sound, scary, wow, love, hard, man \\
        16 & go, time, back, take, need, way, lol, happen, watch, long \\
        17 & fuck, na, shit, gon, holy, man, wan, oh, sick, dude \\
        18 & better, work, great, train, luke, way, deepfakes, job, mean, technology \\
    \end{tabular}
    \caption{}
    \label{tab:conversation20212}
\end{table*}


\subsection{Sample raw data}

10 samples were obtained to make sense of data when grouping similar posts and deriving implications. Table~\ref{tab:topic_sample_2018} shows a sample (raw data) obtained from topic 1 in 2018. Similarly, we obtained from each topic in each year.

\begin{table*}[]
    \centering
    \begin{tabular}{rl}
        \# & Terms \\
        \hline
        01 & everyone use deepfake right \\
        02 & nzz deepfake auch falsch ist ist irgendwie wahr nur anders digital \\
        03 & collection kind deepfake \\
        04 & deepfake anna \\
        05 & tiffany deepfake \\
        06 & hayley williams deepfake \\
        07 & anyone wan na make chrissy costanza deepfake piper perri feel would great use \\
        08 & going abroad deepfake masquerade \\
        09 & amount new deepfake websites last hours amaze \\
        10 & tensor flow machinelearning deepfake
    \end{tabular}
    \caption{}
    \label{tab:topic_sample_2018}
\end{table*}

\end{document}